\documentclass{ws-ijmpd}
\usepackage[super]{cite}
\usepackage{xcolor}
\usepackage[verbose,hypertexnames=false]{hyperref}
\usepackage{graphicx}
\usepackage[numbers]{natbib}
\usepackage{subfigure}
\hypersetup{colorlinks=false,allbordercolors=blue,pdfborderstyle={/S/U/W 1}}

\begin{document}

\markboth{Pravin Kumar Natwariya}
{Axion-Photon Conversion In Magnetized Universe}

%
\catchline{}{}{}{}{}
%

\title{Axion-Photon Conversion In Magnetized Universe: Impact On The Global 21-cm Signal}

\author{Pravin Kumar Natwariya}
\address{School of Fundamental Physics and Mathematical Sciences, Hangzhou
Institute for Advanced Study, University of Chinese Academy of Sciences (HIAS-UCAS), Hangzhou, 310 024, China}
\address{University of Chinese Academy of Sciences, Beijing, 100 190, China}
\address{International Centre for Theoretical Physics Asia-Pacific (ICTP-AP), Beijing, 100 190, China}
\author{Vivekanand Mohapatra}
\address{Department of Physics, National Institute of Technology Meghalaya, Cherrapunji, Meghalaya 793 108, India}
\author{Hriditi Howlader}
\address{Department of Physics, National Institute of Technology Meghalaya, Cherrapunji, Meghalaya 793 108, India}

\maketitle

\begin{history}
\received{(Day Month Year)}
\revised{(Day Month Year)}
\accepted{(Day Month Year)}
\published{(Day Month Year)}
\end{history}

\begin{abstract}

The reported anomalous global 21-cm signal $(T_{21})$ from the cosmic dawn era by Experiment to Detect the Global Epoch of Reionisation Signature (EDGES) could hint towards new physics beyond the standard model. The resonant conversion of the axion-like particles (ALPs) into photons in the presence of primordial magnetic fields (PMFs) could be a viable solution. However, the strength of the PMFs can change over the time as they can decay by ambipolar diffusion and turbulent decay. Consequently, PMFs can dissipate their energy into the intergalactic medium (IGM), which can alter the global 21-cm signal. We simultaneously consider both magnetic heating of IGM and resonant conversion of ALPs to derive physically motivated upper bounds on the coupling strength $(g_{a\gamma})$ and magnetic field strength $(B_n)$. Our findings report that, for $B_n= 0.1\,\rm nG$, $g_{a\gamma}B_n\lesssim (3.6\times 10^{-4}-3\times 10^{-3})$ is required to recover standard $T_{21}=-156\,\rm mK$, while a deeper absorption of $-500$ mK pushes the upper bound to $g_{a\gamma}B_n\lesssim (6.5\times 10^{-4}-5.7\times 10^{-3})$.

\end{abstract}


\section{Introduction}\label{sec1}

Despite the remarkable success of the Standard Model (SM) of particle physics, there are several compelling reasons to consider the possibility of new physics beyond the Standard Model (BSM). Although it requires further confirmation, the recent EDGES observation of the Rayleigh-Jeans tail of the cosmic microwave background (CMB) may offer another hint of BSM physics \cite{Bowman:2018yin}. The detection reported an anomalous absorption depth in the global 21-cm signal of $-0.5_{-0.5}^{+0.2}~\rm K$ centred at $78~\rm MHz~(z\sim 17.2)$ which is nearly twice the $\Lambda\rm CDM$ model predicts \cite{Bowman:2018yin}. The observation suggests a cooler IGM temperature or a hotter CMB in this frequency. The former scenario has been investigated by invoking dark matter-baryon scattering \cite{Barkana:2018qrx, Fialkov:2018xre}, whereas the latter scenario requires the production of extra radio photons in the 21-cm regime of the CMB \cite{Lawson:2012zu, Feng:2018rje, Pospelov:2018kdh, Moroi:2018vci, Lawson:2018qkc, Choi:2019jwx}. Although SARAS 3 has rejected the entire signal with $95\%$ confidence level \cite{Singh:2021mxo}, future experiments such as HERA \cite{DeBoer:2016tnn}, REACH \cite{deLeraAcedo:2022kiu}, MWA-II \cite{Tiwari:2023wzg}, and MIST \cite{Monsalve:2023lvo} may resolve this tension soon.

The latter scenario can be obtained by invoking the resonant conversion of dark radiation into photons. One of the possible candidates for dark radiation is an axion-like particles. Axion-like particles (ALPs) are a potential cold dark matter candidate proposed to solve the strong CP problem in quantum chromodynamics (QCD) \cite{sikivie2010dark}. The Peccei-Quinn mechanism introduces a new global $\,\rm U(1)$ symmetry, called Peccei-Quinn (PQ) symmetry, by spontaneous breaking, resulting in a new pseudoscalar particle which is known as axion \cite{Vysotsky:1978dc, Berezhiani:1989fu, Berezhiani:1992rk, Sakharov:1994id, Sakharov:1996xg, ishida2022peccei, Guo:2023hyp}. On introducing the axion field, one can dynamically cancel the CP-violating term $(\theta_{\,\rm QCD} = 0)$, resolving the strong CP problem. The mass of ALPs is smaller than that of standard model particles and can arise through different mechanisms, like PQ symmetry breaking mechanisms \cite{ishida2022peccei}. The number density spectrum of dark radiation can be of several orders of magnitude greater than CMB; therefore, a minute conversion probability can produce excess radio radiation that can explain the EDGES anomaly \cite{Pospelov:2018kdh, Moroi:2018vci, Choi:2019jwx}. In a previous study \cite{Moroi:2018vci}, the authors have considered the conversion of ALP to photon in the presence of PMFs to explain the EDGES anomalous absorptional amplitude.

The existence of primordial magnetic field (PMF) has been studied extensively to explain the observation of the magnetic fields on the Mpc scale \cite{2002ARA&A..40..319C, 2018MNRAS.479..776V, Neronov:2010gir, Vovk:2011aa}. However, the origin and evolution of large-scale magnetic fields lack a comprehensive understanding. As a result, numerous studies have explored possible mechanisms for their generation, particularly focusing on scenarios in the early universe involving various physical processes. For instance, the generation of primordial magnetic fields (PMFs) can be seeded from the inflationary epoch \cite{1988PhRvD..37.2743T, 1992ApJ...391L...1R}, due to topological defects \cite{1997MNRAS.287....1S}, preheating \cite{2001PhRvD..63j3515B}, or from the Harrison mechanism \cite{2018CQGra..35o4001H}. The specific generation mechanism determines the resulting spectrum of the PMF.

In this work, we consider ALPs as dark radiation that can convert into photons in the presence of primordial magnetic fields. PMFs can dissipate energy into the primordial plasma, increasing the IGM temperature and hence reducing absorptional amplitude of the global 21-cm signal \cite{2005MNRAS.356..778S, Chluba:2015lpa, Kunze:2014, 2019MNRAS.488.2001M, Mohapatra:2024djd}. In the post-recombination era, PMFs can dissipate energy into the intergalactic medium via ambipolar diffusion \cite{2005MNRAS.356..778S} and turbulent decay \cite{2005MNRAS.356..778S}. We simultaneously consider magnetic heating of the IGM and the production of nonthermal radio photons from ALP-photon conversion. On this ground, we derive a physically motivated upper bound on the ALP-photon coupling coefficient by retrieving both the standard and EDGES reported amplitude of the global 21-cm signal.

In the $\Lambda\rm CDM$ framework, the features of the cosmic dawn signal depend primarily on the star formation. The first luminous object in our universe may have formed around $50\times 10^6$ years after the Big Bang \cite{Furlanetto:2006jb, Pritchard:2011xb}. The universe was predominantly filled with neutral hydrogen atoms, with some residual free electrons and protons during this era--- known as the cosmic dawn. The 21-cm line, originating from the hyperfine transition between singlet $(\rm F = 0)$ and triplet $(\rm F = 1)$ state of the ground state of neutral hydrogen, is a treasure trove in the era of precision cosmology. Observations of the 21-cm signal from the cosmic dawn can provide valuable insights not only into the formation of the first stars but also into possible signatures of exotic physics, including the nature of dark matter.

The global (sky-averaged) 21-cm brightness temperature relative to CMB, $T_{21}$, is expressed as \cite{Furlanetto:2006jb, Pritchard:2011xb}
\begin{equation}
T_{21} \approx 27\,(1-x_e)\left(\frac{\Omega_{b}h^{2}}{0.02}\right)\left(\frac{1+z}{10}\frac{0.15}{\Omega_{m}h^{2}}\right)^{1/2}\left(1-\frac{T_{\gamma}}{T_s}\right)\,\rm mK,
\label{eq:T21}
\end{equation}

where, $x_e$ is the ionization fraction, $\Omega_{b}$ and $\Omega_{m}$ are the baryon and matter density parameters, respectively. $z$ represents redshift and $h$ is the Hubble parameter in the units of $100\,\rm Km/s/Mpc$ \cite{ade2016planck}. $T_{\gamma} = T_{\gamma, 0}(1+z)$ is the cosmic microwave background (CMB) temperature, with $T_{\gamma,0} = 2.725~\rm K$ being the present-day temperature. $T_s$ represents the spin temperature that determines the relative population density of neutral hydrogen atoms. In the $\Lambda\rm CDM$ framework, the spin temperature is governed by three mechanisms: 1) The scattering between CMB photons and ionized component of the IGM \cite{Venumadhav:2018uwn}, 2) The collisions between ionized-ionized and ionized-neutral components in the IGM via spin-exchange transition \cite{1958PIRE...46..240F}, and 3) The Lyman alpha photons originating from the first star via Wouthysen-Field effect \cite{1952AJ.....57R..31W, 1959ApJ...129..536F}. The spin temperature is given by \cite{Furlanetto:2006jb, Pritchard:2011xb}

\begin{equation}
    T_s^{-1}= \frac{x_{\gamma}T_{\gamma}^{-1}+(x_{c}+ x_{\alpha})T_{g}^{-1} }{x_{\gamma}+ x_{c} + x_{\alpha}}~,
    \label{Ts}
\end{equation}
where $T_g$ represents the intergalactic medium (IGM) temperature. $x_\gamma$, $x_c$, and $x_{\alpha}$ represent the CMB, collisional, and Lyman alpha coupling coefficients, respectively. Given the low optical depth for 21-cm photons, $x_{\gamma}\approx 1$. From Eq. \eqref{eq:T21}, we see that one expects an absorption signal for $T_s<T_{\gamma}$. The IGM and CMB temperatures evolve as $(1+z)^2$ and $(1+z)$, respectively, due to the expansion of the universe. CMB and IGM share the same temperature at $z>200$ due to the efficient inverse Compton scattering; however, at $z\lesssim200$ the Compton scattering rate becomes insignificant compared to the Hubble rate, resulting in their independent thermal evolution. Electromagnetic radiations, such as Lyman alpha $(\rm Ly\alpha)$ and X-rays originating from the astrophysical sources, can heat and ionize the IGM. Additionally, the Ly$\alpha$ photons couple the $T_s$ and $T_g$ via the WF effect, resulting in an absorption signal at redshifts $12\lesssim z\lesssim 30$.

\section{Axion-photon conversion and its effect on CMB}
\label{axion-photon}

In this section, we discuss the resonant conversion of axion-like particles (ALPs) and photons in the presence of magnetic fields. The ALPs can be considered to have a coupling with photon as $\mathcal{L} = 1/4\,g_{a\gamma}a\, F_{\mu\nu}\tilde{F}^{\mu\nu}$, where $F_{\mu\nu}$ and $\tilde{F}_{\mu\nu}$ represent the ordinary electromagnetic gauge field and its dual field strength, respectively. $g_{a\gamma}$ and $a$ represent the ALP-photon coupling strength and axion field, respectively. In the presence of background magnetic fields, $g_{a\gamma}$ can give an effective mixing between ALP and photon \cite{Raffelt:1987im, Pospelov:2018kdh, Moroi:2018vci, Choi:2019jwx}. In this work, we consider the background magnetic field to be primordial magnetic fields (PMF) with present-day strength and spectral index as $B_n$ and  $n_B$, respectively. Now, consider ALPs with energy $E_a$ traversing through magnetic fields, with $\bf{B}$ strength, perpendicular to their momentum. For a resonant conversion to occur, the mixing angle $(\theta)$ between ALP and photon should be maximum, which is given by \cite{Mirizzi:2009nq, Moroi:2018vci},

\begin{equation}
    \sin^2(2\theta) = \frac{4g_{a\gamma}^2(E_a\bf{B})^2}{ \left(\omega_p^2 - m_a^2\right)^2+4g_{a\gamma}^2(E_a\mathbf{B})^2}\, ,
    \label{Eq:mixing_angle}
\end{equation}

where $\omega_p$ and $m_a$ represent the effective photon mass and mass of ALP, respectively. In the post-recombination era, $\omega_p$ evolves due to evolution of the ionization fraction $(x_e)$, which is given by \cite{Moroi:2018vci}

\begin{equation}
    \omega_p^2 \simeq \left(1.6\times 10^{-14}\,\rm eV\right)^2 (1+z)^3x_e(z)\, .
    \label{eq:mass_alp}
\end{equation}
From Eq. \eqref{Eq:mixing_angle}, we can see that the mixing angle attains a maximum value when $\omega_p^2 = m_a^2$ \cite{Yanagida:1987nf, Mirizzi:2009nq, Moroi:2018vci}. As $\omega_p^2\propto x_e$, its value decreases over redshift due to the recombination and expansion of the universe. Consequently, the resonant conversion of ALP, of mass $m_a$, to a photon can be uniquely determined from the evolution of $x_e(z)$ at a particular resonance redshift, $z = z_{\rm res}$.

\begin{figure}[htbp]
    \centering
    \includegraphics[width=\linewidth]{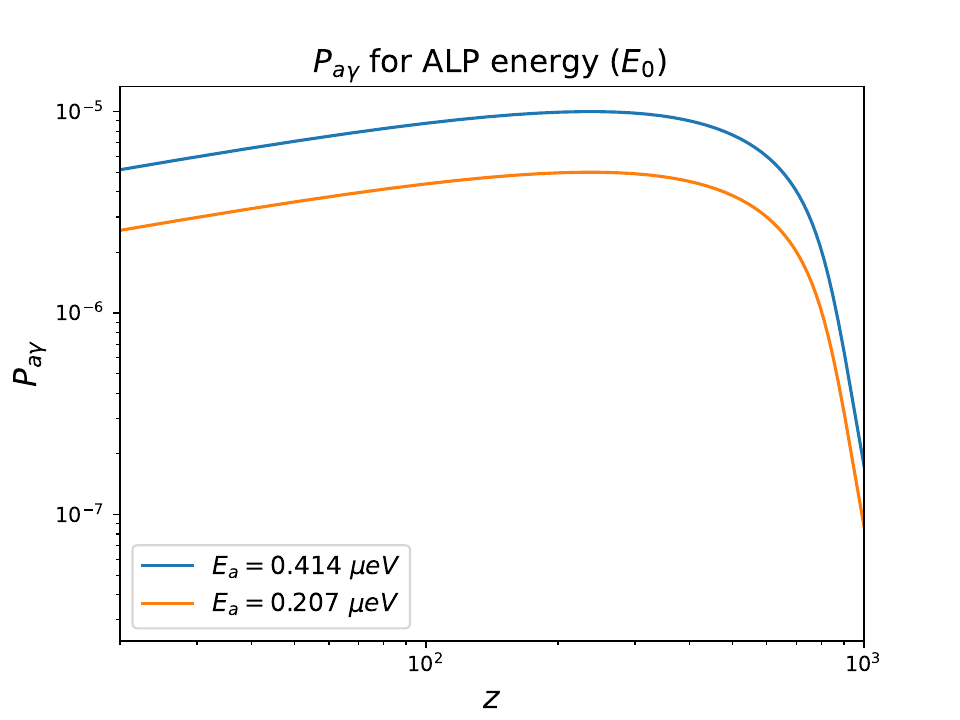}
    \caption{Represents the evolution of conversion probability $(P_{a\gamma}) $ over redshifts for $g_{a\gamma}B_n = 10^{-12}\,\rm nG\, GeV^{-1}$ and energies $E_a = 0.414\,\mu\rm eV$ (blue) and $E_a = 0.207\,\mu\rm eV$ (orange).}
    \label{fig:prob_conversion}
\end{figure}

We considered the evolution of $x_e$ at redshifts $\lesssim 1100$, as the resonant photons produced at $z>1100$ (pre-recombination era) can get absorbed by the primordial plasma efficiently \cite{Mirizzi:2009nq, Chluba:2015hma, Pospelov:2018kdh, Moroi:2018vci, Choi:2019jwx}. Using $\texttt{Recfast++}$ \cite{10.1111/j.1365-2966.2010.16940.x, 10.1111/j.1365-2966.2010.17940.x}, we find that the mass of ALP resonantly converting to photons at redshifts $20\lesssim z\lesssim 1100$ should be in the range $\sim\left(10^{-14}-10^{-10}\right)\,\rm eV$ (see Eq. \ref{eq:mass_alp}).
Now, the full-sky and polarization averaged ALP-photon conversion probability $(P_{a\gamma})$ is given by \cite{Mirizzi:2009nq, Moroi:2018vci}

\begin{equation}
    P_{a\gamma} \simeq \frac{\pi rg_{a\gamma}^2 E_a}{m_a^2}\frac{\langle \mathbf{B}^ 2\rangle}{3}\Bigg{|}_{z = z_{\rm res}}~,
    \label{eq:coversion_probability}
\end{equation}
where $r \simeq 3H$ in the post-recombination era \cite{Moroi:2018vci}, and $\mathbf{B} = B_n(1+z)^2$ \cite{Durrer:2013pga}. In Fig. \eqref{fig:prob_conversion}, we present the evolution of $P_{a\gamma}$ for ALPs with energies $0.414\,\mu \rm eV$ (blue) and $0.207\,\mu \rm eV$ (orange) at redshifts $20-1000$. The energies $0.414\,\mu \rm eV$ and $0.207\,\mu \rm eV$ correspond to the frequencies $100\,\rm MHz$ and $50\,\rm MHz$ at the present epoch, respectively, representing redshifted 21-cm photons from redshifts 13.2 and 27.4. Here, we have considered $g_{a\gamma}B_n = 10^{-12}\,\rm GeV^{-1}\,nG$, for illustration. We can observe a sharp rise in the $P_{a\gamma}$ at redshifts $250\lesssim z\lesssim 1000$ due to the fall in $x_e$ after recombination. At these redshifts, $x_e$ falls rapidly and attains a nearly redshift independent value of $\mathcal{O}\left(10^{-4}\right)$ at $z\lesssim 200$ in $\Lambda\rm CDM$ framework. Therefore, $P_{a\gamma}$ attains a maximum value and then falls as $ (1+z)$. 

Since the conversion occurs in the presence of primordial magnetic fields (PMFs), we adopt the following approximations, as also implemented in Ref. \cite{Moroi:2018vci}. First, we consider primordial magnetic fields with coherence length of $1\,\rm Mpc$ such that the oscillation length of ALP-photon $<1\,\rm Mpc$ at all the redshifts under consideration (see Eq. 9 of Ref. \cite{Moroi:2018vci}). Furthermore, the oscillation length must also be shorter than the mean free path of photons, which scales as $ x_e^{-1}$ \cite{Moroi:2018vci}. For the redshifts $20\lesssim z \lesssim 1000$, both these conditions remain satisfied. In the next section, we will show that the standard evolution $x_e$ remains unaffected even when we consider the magnetic heating of the IGM due to PMFs in the post-recombination era.

\begin{figure}[htbp]
    \centering
    \includegraphics[width=\linewidth]{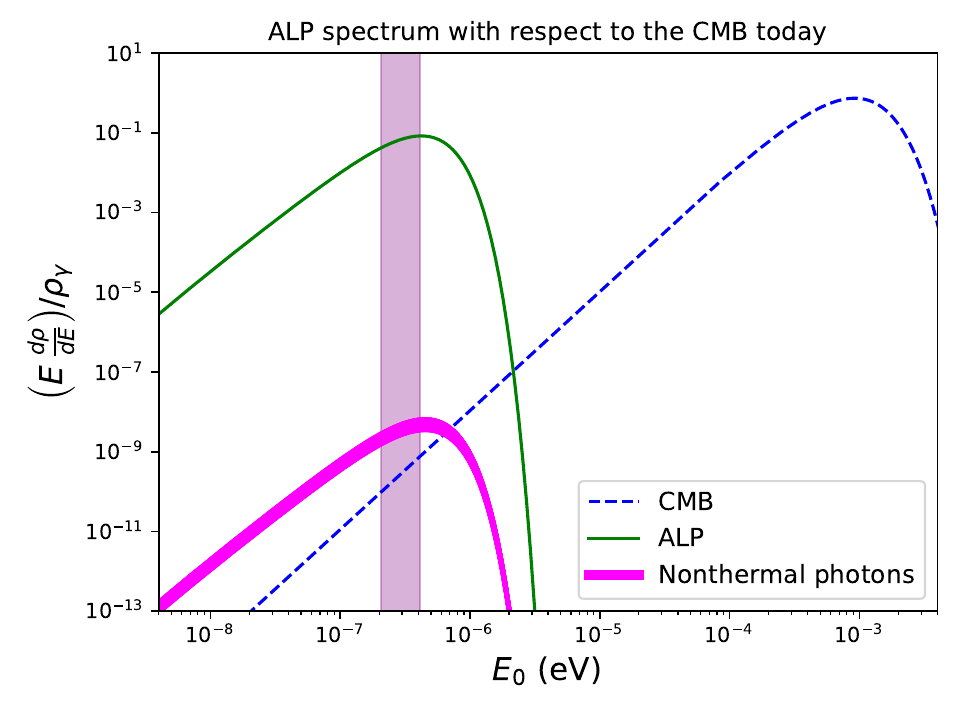}
    \caption{The ratio of energy density spectrum $(Ed\rho/dE)$ and energy density of CMB $(\rho_{\gamma})$. The blue dashed and green solid lines illustrate the ratio of present-day CMB $(Ed\rho_{\gamma}/dE)$ and ALP $(Ed\rho_a/dE)$ energy density spectrum, respectively, with respect to $\rho_{\gamma}$. The magenta band shows nonthermal photons for ALP energies $E_a$ in the range $(50-100)\,\rm MHz$. The vertical purple band shows $(50-100)\,\rm MHz$ frequency band.}
    \label{fig:ALP_spectrum}
\end{figure}

We can now calculate the number density of photons produced on the resonant conversion of ALPs to photons with respect to the CMB spectrum. Let us assume that some scalar field decays to produce ALPs $(\phi\to aa)$, such that the ALPs interact weakly enough to traverse as free particles. Now the redshifted ALP spectrum (number density) with energy $E_a$ can be defined as \cite{Pospelov:2018kdh, Moroi:2018vci, Choi:2019jwx}

\begin{equation}
    \frac{d\rho_a(z)}{dE_a} = \frac{2n_{\phi}(z_d) (1+z)^3\Gamma_{\phi}}{H(z_d)(1+z_d)^3}\, ,
    \label{eq:ALP_spectrum}
\end{equation}
where $n_{\phi}$ and $\Gamma_{\phi}$ represent the number of density and decay rate of $\phi$, respectively. The redshift $z_d$ represents the time of ALPs production. It can be expressed as $1+z_d = (1+z){m_{\phi}}/{2E_a}$, where $m_{\phi}$ is the mass of $\phi$. Following Ref. \cite{Moroi:2018vci}, we take $n_{\phi} = Y_{\phi}s(z)\exp\left[-\Gamma_{\phi}t(z)\right]$, where $Y_{\phi} = 22$ parameterizes $n_{\phi}$, $m_{\phi} = 40\, \mu\rm eV$, and $\Gamma_{\phi} = 10^{-15}\,\rm sec^{-1}$. $s(z)$ represents the entropy density that can be expressed as $s(z_d) = g_*\left[2\pi^2T_{\gamma}^3(z_d)/45\right]$ at redshift $z_d$ --- here $g_* = 3.36$ is the effective degree of freedom for $T_{\gamma}\lesssim \rm eV$ \cite{Piattella:2018hvi}. It is worth noting that the fiducial values of $m_{\phi}$, $Y_{\phi}$, and $\Gamma_{\phi}$ have been considered such that the ALP energy density spectrum will be maximum in the energy (or frequency) band of the EDGES detection. One can certainly vary these values to obtain different spectra. We have not addressed the origin of the scalar field $\phi$ yet. The detailed discussion on the mechanism of its origin is beyond the scope of this paper. However, one can adopt the mechanism proposed in the article \cite{Moroi:2018vci}, where the authors utilized saxions decaying into ALPs. Alternatively, authors in the article \cite{Choi:1996vz} have proposed decay of flaton into ALPs. In this work, we directly use the result obtained in the article \cite{Moroi:2018vci}.

We can now calculate the spectrum of nonthermal photons produced from the conversion of ALP to photons as

\begin{equation}
    \frac{d\rho_{\gamma}}{dE_0} = \frac{d\rho_a}{dE_a}\Bigg{|}_{z = 0}\times P_{a\gamma}(z_{\rm res})
    \label{eq:conversion_spectrum}
\end{equation}
evaluated at resonance redshifts $z_{\rm res}$. In Fig. \eqref{fig:ALP_spectrum}, we represent the ratio of present-day energy density spectra $(E\, d\rho/dE)$ to the CMB energy density $(\rho_{\gamma})$ for energies $E_0$ that can translate to the frequency range $1\,\rm MHz-10^3\,\rm GHz$. The blue dashed and green solid lines represent the ratio for CMB photons $(E\, d\rho_{\gamma}/dE)$ and ALPs $(E\, d\rho_a/dE)$ with $\rho_{\gamma} = (\pi^2/15)\, T_{\gamma,0}$. The magenta solid band presents the ratio of energy density spectra of nonthermal photons obtained using the above equation within the frequency range $(50-100)\,\rm MHz$. Here, we vary the energy $(E_a)$ of the ALPs in $P_{a\gamma}(z_{\rm res})$ from $50\,\rm MHz-100\,\rm MHz$, which corresponds to the EDGES frequency band--- shown in the vertical purple band. As $P_{a\gamma}$ varies for different $E_a$ values, we can limit the conversion probability to obtain the desired present-day nonthermal photons relative to the CMB. These nonthermal photons obtained today can explain the excess radio background during the cosmic dawn era reported by EDGES \cite{Bowman:2018yin}. 

The fraction of the CMB energy $(\mathcal{F}_{\gamma})$ that lies in the EDGES frequency band $(50-100\,\rm MHz)$  today, can be calculated by integrating $(d\rho_{\gamma}/dE_0) /\rho_{\gamma}$ in the energy range $(0.207-0.414)\,\rm \mu eV$. We find the fraction to be $\mathcal{F}_{\gamma} = 2.46\times 10^{-10}$. Similarly, the fraction of ALP spectrum $(\mathcal{F}_{a})$ in the EDGES frequency band can be obtained by integrating $(d\rho_{a}/dE_0) /\rho_{a}$, where the ALP spectrum $(\rho_a)$ and $(d\rho_{a}/dE_0)$ are obtained from Eq. \eqref{eq:ALP_spectrum}. We find the ALP spectrum fraction to be $\mathcal{F}_a = 0.4$. The reported absorption amplitute of $-0.5_{-0.5}^{+0.2}~\rm K$ centred at frequency $78~ \rm MHz~(0.324~\mu\rm eV)$ by EDGES, suggests $\alpha = T_{\gamma}^{\rm EDGES}/T_{\gamma}^{\rm CMB}$ amounts of excess photons at $\nu = 78\,\rm MHz$ today. The term $\alpha$ can be interpreted conventionally as follows: $\alpha = 1$ represents the reference 21-cm signal, i.e., the temperature term in the denominator of $\alpha$; whereas for $\alpha>1$ one expects an excess absorption depth at $z = 17.2 ~(\nu = 78\,\rm MHz)$ compared to the reference signal. The required nonthermal photons, defined as $\rho_a\mathcal{F}_aP_{a\gamma}(0.324~\mu\rm eV)$, should be equal to $\alpha\rho_{\gamma}\mathcal{F}_{\gamma}$ to explain the excess absorption.
In the next section, we will formulate the evolution of the global 21-cm signal in the presence of primordial magnetic fields, and in Sec. \eqref{sec: Result} we will obtain bounds on $P_{a\gamma}$ for the different $T_{21}$ absorption amplitudes at $\nu = 78~\rm MHz$.

\section{Evolution of IGM in the presence of PMF}\label{sec: thermal_evolution}

In the post-recombination era, primordial magnetic fields can dissipate energy into the intergalactic medium (IGM) via ambipolar diffusion and turbulent decay \cite{2005MNRAS.356..778S, Kunze:2014, 2019MNRAS.488.2001M}. In the pre-recombination era, turbulent motions in the plasma were heavily damped due to large radiative viscosity. However, after recombination as the universe became neutral, these dampings lowered, leading to transfer of magnetic energy below the Jeans length scale--- known as turbulent decay \cite{2005MNRAS.356..778S}. Additionally, the magnetic fields can accelerate the ionized component of the IGM due to the Lorentz force. These accelerated components then collide efficiently with the neutral components of the IGM, dissipating the magnetic energy--- known as ambipolar diffusion \cite{2005MNRAS.356..778S}. The energy dissipation rate depends on the spectrum of PMF, which depends on its origin. We consider a statistically isotropic and homogeneous Gaussian random field with zero helicity. The statistical properties of the field can be completely determined from its power spectrum \cite{landau2013statistical}. We consider a power law-like power spectrum which is given by \cite{2005MNRAS.356..778S, Kunze:2014, 2019MNRAS.488.2001M, Natwariya:2022byx},

\begin{equation}
	P_B(k) = \frac{(2\pi)^{n_B+5}\,B_n^2}{\Gamma\left[\left(n_B + 3\right)/2\right] k_n^{3}}\left(\frac{k}{k_n}\right)^{n_B}\, ,
	\label{power_spectrum}
\end{equation}
where $B_n$, $k_n= 1\,\rm Mpc^{-1}$, and $n_B$ represent the present-day magnetic field amplitude, a smoothing scale, and spectral index, respectively. Before recombination, radiative viscosity dampens the PMFs on length scales smaller than a cutoff scale $k_c^{-1}$. However, $k_c$ can change over time as the universe evolves \cite{PhysRevD.58.083502}. We consider a sharp cutoff of $P(k)$ for all $k>k_c$. The evolution of $k_c$ with redshift can be expressed as $k_c = g(z)\,k_{c,i}$. Here, $g(1088) = 1$ \cite{2019MNRAS.488.2001M}, and $k_{c,i}$ is the cutoff scale at $z = 1088$ while $g(z)$ determines the redshift evolution of $k_c$.

The evolution of IGM temperature $(T_g)$ with redshift in the presence of primordial magnetic fields can be written as \cite{2005MNRAS.356..778S, Schleicher:2008aa, Chluba:2015lpa, 2019MNRAS.488.2001M, natwariya2021constraint, PhysRevD.98.103529, Mohapatra:2024djd},

\begin{alignat}{2}
	\frac{dT_g}{dz} =  &\frac{2\,T_g}{(1+z)} + \frac{\Gamma_c}{(1+z)H(z)} \left(T_g - T_{\gamma}\right)  \nonumber \\ & -\frac{2}{3(1+z)H(z)n_{\rm tot}} \left(\Gamma_{\rm AD} + \Gamma_{\rm TD}\right),
    \label{Eq:thermal_evolution}
\end{alignat}

where $\Gamma_c = \frac{8 n_e\sigma_T a_rT_{\gamma}^4 (z)}{3m_e n_{\text{tot}}}$ is the Compton scattering rate. Here, $a_r$ and $n_{\rm tot} = n_H(1+f_{He}+x_e)$ are the radiation density constant and total number density of gas, respectively. $f_{He}$, $m_e$, and $\sigma_T$ represent helium fraction, electron's rest mass, and Thomson scattering cross-section, respectively. The third term on the right-hand side of Eq. \eqref{Eq:thermal_evolution} represents energy dissipation from PMFs. $\Gamma_{\rm AD}$ and $\Gamma_{\rm TD}$ represent the volumetric energy dissipation rate via ambipolar diffusion (AD) and turbulent decay (TD), respectively \cite{2005MNRAS.356..778S}. The terms $\Gamma_{\rm TD}$ is given by \cite{2005MNRAS.356..778S, Kunze:2014, 2019MNRAS.488.2001M, Mohapatra:2024djd},

\begin{equation}
	\Gamma_{\rm TD} = \frac{3\,\mathrm{m}\,E_B\,H(z)}{2}\frac{[\ln(1+t_d/t_{\rm rec})]^{\mathrm{m}}}{\left[\ln(1+t_d/t_{\rm rec}) + \ln(t/t_{\rm rec})\right]^{1+\mathrm{m}}}\, ,
	\label{turb_decay_eqn}
\end{equation}

where, $\mathrm{m} = 2(n_B+3)/(n_B+5)$, $t_d = 1/\left(k_cV_a\right)$ for $n_B>-3.0$ represents Alfvén time scale for the cutoff scale \cite{2005MNRAS.356..778S}, $t_{\rm rec} = 2/(3H)$ is recombination time-period in the matter-dominated universe, and $E_B = B^2/8\pi$ is the magnetic energy density. The energy dissipation via AD is given by

\begin{equation}
	\Gamma_{\rm AD} = \frac{2(1 - x_e)/x_e}{32\pi^2\zeta_e(m_Hn_H)^2}\,\bigg{|}\left(\vec{\nabla}\times \vec{B}\right)\times \vec{B}\bigg{|}^2\, ,
	\label{ambipolar_rate_eqn}
\end{equation}

where $\zeta_e$, $m_H$, $n_H$ represent the ion-neutral coupling coefficient, mass and number density of hydrogen atoms \cite{2005MNRAS.356..778S, 2019MNRAS.488.2001M}. We note that the derivative operator is taken with respect to the proper coordinate and $B \sim B_n(1+z)^2$. The evolution of $E_B$ with redshift in the presence of energy dissipation from Eqs. \eqref{turb_decay_eqn} and \eqref{ambipolar_rate_eqn} can be expressed as \cite{2005MNRAS.356..778S, Kunze:2014, 2019MNRAS.488.2001M},
\begin{equation}
	\frac{dE_B}{dz} = \frac{4\,E_B}{1+z} + \frac{\left(\Gamma_{\rm TD} + \Gamma_{\rm AD}\right)}{(1+z)H(z)}.
	\label{mag_enrg_cons}
\end{equation}

Here, the first term corresponds to the redshifting of magnetic energy due to the adiabatic expansion of the universe. In contrast, the second term corresponds to the magnetic energy lost via the AD and TD processes.

    

\begin{figure*}
    \centering
    \subfigure[]{\includegraphics[width=0.45\textwidth]{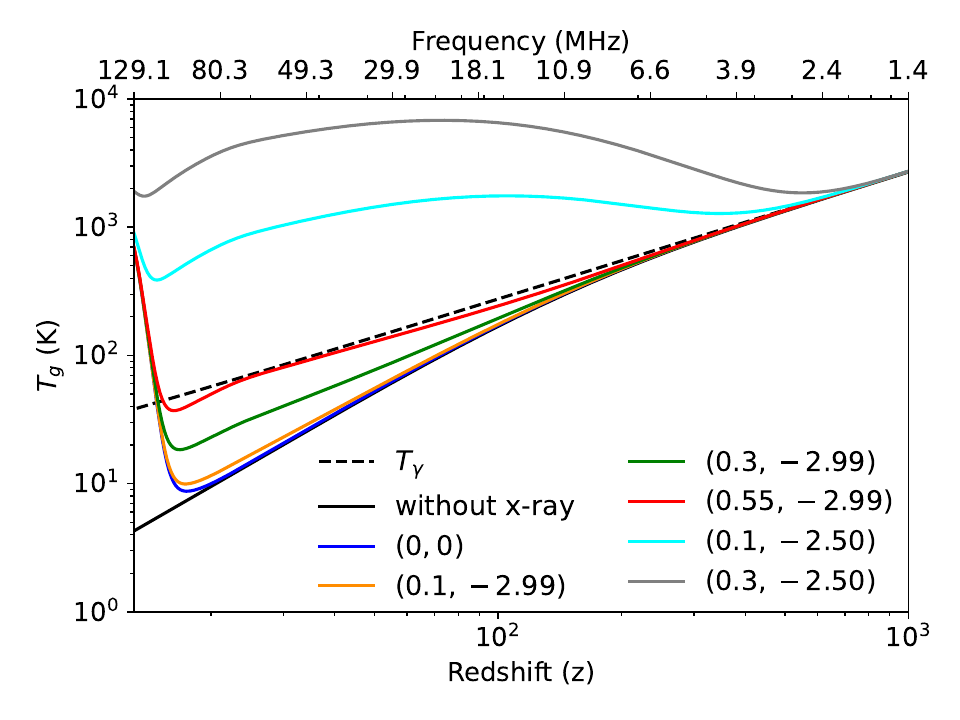}
        \label{fig:Gas_temp_tanh_PMF}}
    \hfill
    \subfigure[]{\includegraphics[width=0.45\textwidth]{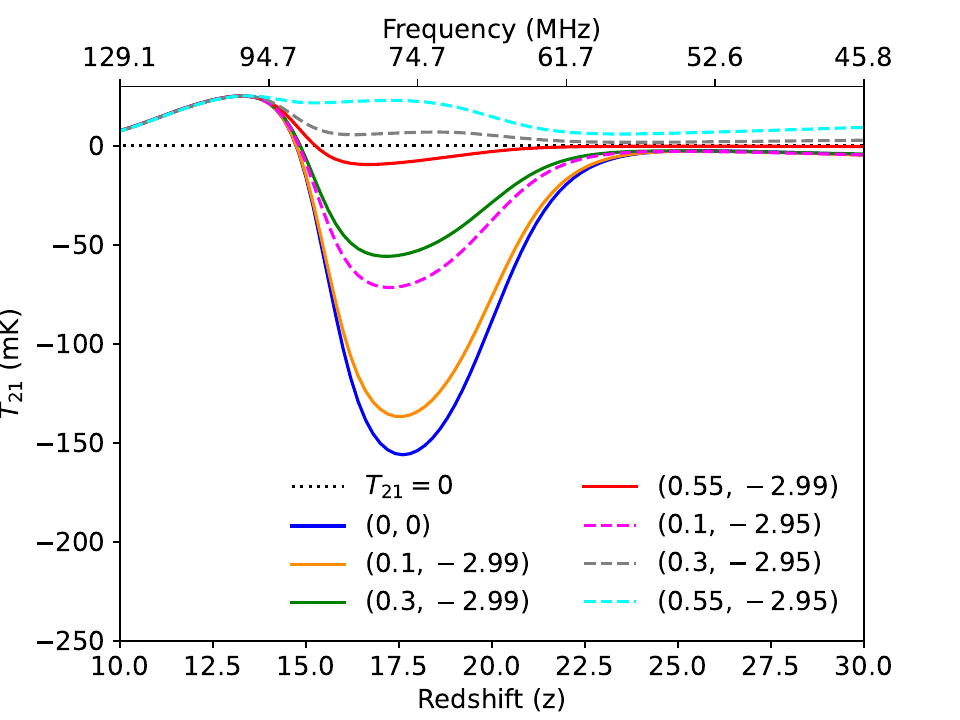}
        \label{fig:T21_plot}}
    
    \caption{ (a) Thermal evolution of the IGM with primordial magnetic fields. The black dashed and solid lines show the CMB temperature and $T_g$ in $\Lambda\rm CDM$ without star formation. The blue solid line includes X-ray heating. Yellow, green, and red solid lines show $T_g$ for PMFs with $n_B = -2.99$ and $B_n = 0.1, 0.3, 0.55\,\rm nG$, respectively. Cyan and grey lines correspond to $(B_n/\mathrm{nG}, n_B) = (0.1,-2.50)$ and $(0.3, -2.50)$.
        (b) Evolution of global 21-cm signals as a function of redshift in the presence of PMFs. Black dotted and blue solid lines represent $T_{21} = 0$ and the $\Lambda\rm CDM$ signal. Solid and dashed lines show PMF cases with different $(B_n/\mathrm{nG}, n_B)$ values.
    }
    \label{fig:Gas_temp}
\end{figure*}

Primordial magnetic fields cannot ionize the IGM directly. Instead, the dissipated energy can increase the relative velocity between the ionized and neutral components of the IGM. This led to collisional-ionization of the IGM. However, it is exponentially suppressed and efficient only for $T_g\gg 10^4\, \rm K$, which has not been considered in this work \cite{Chluba:2015lpa}. Thus, upon ignoring this effect, we write the ionization evolution of the IGM as \cite{Peebles:1968ja, Seager:1999bc, Chluba:2015lpa},
\begin{alignat}{2}
	\frac{dx_e}{dz} = \frac{1}{(1+z)H(z)}  &\frac{1+ \mathcal{K}_H\Lambda_Hn_H(1-x_e)}{1+ \mathcal{K}_H(\Lambda_H+\beta_H)n_H(1-x_e)} \nonumber \\ & \Big[n_Hx_e^2\alpha_B  - (1-x_e)\beta_Be^{-E_{\alpha}/T_{\gamma}}\Big],
	\label{xe_evolution}
\end{alignat}

where $\alpha_B$ and $\beta_B$ are the case-B recombination and photo-ionization rates, respectively \citep{Seager:1999bc, Seager:1999km, Mitridate:2018iag}. $\mathcal{K}_H = \pi^2(E_{\alpha}^3H)^{-1}$ represents redshifting Ly${\alpha}$ photons, $E_\alpha = 10.2\,\rm eV$ is the first excitation energy of hydrogen atom, and $\Lambda_H = 8.22\,\rm sec^{-1}$ is the 2S-1S level two-photon decay rate in hydrogen atom \cite{PhysRevA.30.1175}. We note that the presence of PMFs does not affect the evolution of $x_e$; therefore, the effective photon mass mentioned in the previous section remains unaltered and follows the $\Lambda\rm CDM$ framework.

\begin{figure*}
\includegraphics[width=\textwidth,height=0.85\textheight, keepaspectratio]{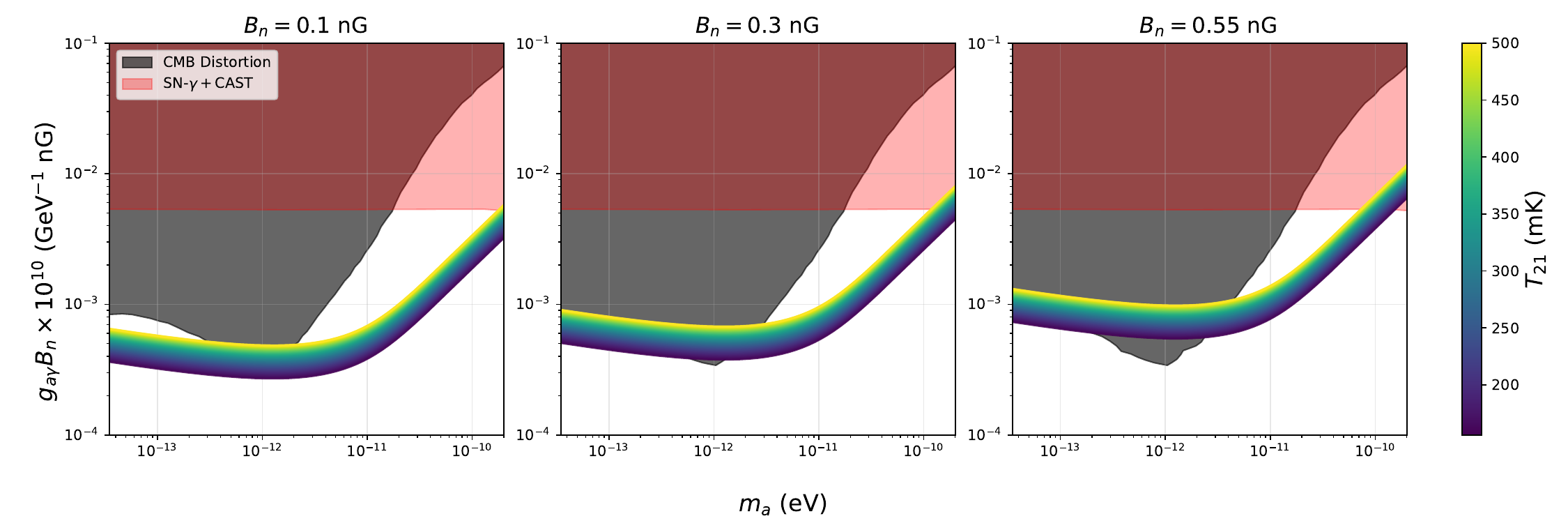}
    \caption{Constraint on $g_{a\gamma}B_n$ and mass of ALP $(m_a)$ in the presence of nearly scale-invariant PMFs with different strengths. The black shaded region represents constraints from spectral distortion \cite{Mirizzi:2009nq}. The red shaded region represents the excluded parameter space from the absence of the $\gamma$-ray burst associated with SN1987A + CAST \cite{Payez:2014xsa}. The colour-coded band presents the variations of $g_{a\gamma}$ with $m_a$ such that the nonthermal resonant photons can produce an absorption amplitude of $T_{21}$ signal at $z\sim 17.2$ in the range $(156-500)\,\rm mk$. Here $T_{21} = -156\,\rm mK$ represents the absorption amplitude in the $\Lambda\rm CDM$ framework.}
    \label{fig:constrain}
\end{figure*}

\section{Results and Discussion}
\label{sec: Result}

In this section, we present the thermal evolution of the IGM and the redshift dependence of the global 21-cm signal. Further, we determine the required axion-photon coupling coefficient to produce sufficient nonthermal radio photons capable of explaining the amplitude of the global 21-cm signal reported by the EDGES collaboration \cite{Bowman:2018yin}. We begin by incorporating X-ray heating of the IGM and the Lyman alpha coupling $(x_{\alpha})$, following the methodology of articles \cite{PhysRevD.98.103529, Mohapatra:2024djd}. These works implement a $tanh$-based prescription to model the X-ray heating term in Eqs. \eqref{Eq:thermal_evolution} and \eqref{xe_evolution} for a standard star formation scenario. We solve Eqs. \eqref{Eq:thermal_evolution}, \eqref{xe_evolution}, and \eqref{mag_enrg_cons} simultaneously, using initial conditions $T_g = 2967.6\text{ K}$, and $x_e = 0.1315$ at redshift $z = 1088$ from \texttt{Recfast++} \cite{10.1111/j.1365-2966.2010.16940.x, 10.1111/j.1365-2966.2010.17940.x}, and $g(1088) = 1$ \cite{2019MNRAS.488.2001M}. For a detailed discussion, we request readers to refer to the article \cite{Mohapatra:2024djd}. 

In Fig. \ref{fig:Gas_temp_tanh_PMF}, the black dashed and solid lines represent the evolution of CMB and IGM temperature, respectively, in the absence of X-ray and magnetic heating in the $\Lambda\rm CDM$ framework. The blue solid line shows a sharp increase in $T_g$ at $z\lesssim 20$ due to X-ray heating. We then include PMFs heating with different field strengths and spectral indices. For example, the yellow, green, and red solid lines represent $T_g$ evolution in the presence of nearly scale-invariant magnetic fields with spectral index $n_B = -2.99$ and present-day field strength $B_n = 0.1, 0.3, 0.55\,\rm nG$, respectively. We note that, on increasing $B_n$, $T_g$ increases significantly and for $(B_n/\mathrm {nG}, n_B) = (0.55, -2.99)$, the IGM temperature becomes comparable to $T_{\gamma}$ at certain redshifts. Further, we increase the spectral index to $-2.50$ and calculate $Tg$ for $B_n$ values of $0.1\,\rm nG$ (cyan line) and $0.3\,\rm nG$ (gray line) illustrating that the magnetic heating can be strong enough to raise $T_g$ above $T_{\gamma}$.

In Fig. \ref{fig:T21_plot}, we present the redshift evolution of the global 21-cm signal using Eq. \eqref{eq:T21} by calculating spin temperature using Eq. \eqref{Ts}. The black dotted line corresponds to $T_{21} = 0$, while the blue solid line represents $T_{21}$ in the presence of X-ray heating.
Using the fiducial values of the $tanh$ parameters for X-ray heating and Lyman alpha coupling from article \cite{Mohapatra:2024djd}, we find that $T_{21}(z \simeq 17.2) =-156\,\rm mK$. Including PMFs with $n_B = -2.99$ and varying $B_n = 0.1,0.3,0.55~\rm nG$, we obtain absorption amplitudes of $-132.5~\rm mK$, $-54.6~\rm mK$, and $-9.5~\rm mK$, respectively--- as shown in the yellow, green, and red solid lines. It can be seen that a nearly scale-invariant magnetic field can potentially reduce the absorption amplitude of global 21-cm signals, and could erase it during the cosmic dawn era. To further examine such scenarios, we increased the spectral index to $-2.95$ while keeping the strengths fixed to $0.1, 0.3, 0.55\,\rm nG$--- as shown in the magenta, grey, and cyan dashed lines. We can see that a mild increase in the spectral index can significantly alter $T_{21}$ signals. This signifies that the inclusion of magnetic heating of the IGM is crucial when interpreting the EDGES observation reported excess absorption depth from ALP-photon conversion in the presence of primordial magnetic fields.

Next, we quantify the photon excess fraction ($\alpha$) required to restore the standard $\Lambda\rm CDM$ signal $T_{21}^{\Lambda\rm CDM} = -156~\rm mK$ in the presence of PMF heating. For instance, with $(B_n/\mathrm{nG}, n_B) = (0.1, -2.99)$, we get $T_{21} = -132.5\,\rm mK$, leading to $\alpha = T_{21}^{\rm \Lambda CDM}/T_{21}^{\rm PMF} = 1.18$.
We then calculate $\alpha$ for other $B_n$ values such that $T_{21}^{\Lambda\rm CDM}$ will be retrieved from the nonthermal photons. Similarly, to obtain the EDGES absorption, we considered $T_{21}(z \simeq 17.2)$ equal to $-300\,\rm mK$ and $-500\,\rm mK$, which represent the upper limit and the reported value, respectively. The new values of $\alpha$ can be defined for these $T_{21}$s by replacing $T_{21}^{\Lambda\rm CDM}$ with the desired values.

We then find upper bounds on $g_{a\gamma}$ such that $\mathcal{F}_a\rho_aP_{a\gamma} = \alpha\rho_{\gamma}F_{\gamma}$ over $m_a\sim ( 10^{-14}- 2\times 10^{-10})~ \rm eV$ at a fixed axion energy $E_a = 0.323~\rm\mu eV$ which corresponds to $78\,\rm MHz$. These results are shown in Fig. \eqref{fig:constrain} for $n_B = -2.99$ and $B_n = 0.1, 0.3, 0.55~\rm nG$ from left to right panel, where the color-coded band represents the variation of $T_{21}$ from $-156~\rm mK$ to $-500~\rm mK$. We considered a wide range of $T_{21}$ values to depict various possible scenarios from the standard $\Lambda\rm CDM$ framework to the presence of excess radiation. We restrict our analysis to $n_B = -2.99$ as larger $n_B$ values lead to greater dissipation of magnetic energy, which can erase the $T_{21}$ signal. Furthermore, magnetic fields $B_n\gtrsim 0.55\,\rm nG$ results in heating the IGM $\gtrsim T_{\gamma}$. We find that $g_{a\gamma}$ increases with increasing absorption depth, since a larger $\alpha$ is required, implying a higher conversion probability $P_{a\gamma}$ and hence a stronger coupling. A similar trend can be observed in increasing $B_n$, as the presence of magnetic heating reduces the absorptional amplitude of the $T_{21}$ signal.

Finally, we note that lowering $B_n$ below $0.1~\rm nG$ may reduce the heating, but this simultaneously increases the required $g_{a\gamma}$, due to its $\propto B_n^{-1}$ dependence. Thus, the determination of $g_{a\gamma}$ is a trade-off between the magnetic heating and its strength $B_n$. Our results indicate that to recover $T_{21} = -156~\rm mK$ with $B_n = 0.1~\rm nG$, the coupling must satisfy $g_{a\gamma}B_n\lesssim (3.6\times 10^{-4}-3\times 10^{-3})$, while for the EDGES amplitude of $-500~\rm mK$, the bound relaxes to $g_{a\gamma}B_n\lesssim (6.5\times 10^{-4}-5.7\times 10^{-3})$. For comparison, we present the excluded region from spectral distortion in black color \cite{Mirizzi:2009nq}. Whereas, the red shaded region represents the excluded parameter space from non-observation of $\gamma$-ray burst associated with SN1987A + CAST \cite{Payez:2014xsa}.

\section{Summary and Conclusion}\label{conclusion}

The recently reported anomalous absorption signal by EDGES suggested the possibility of excess radio background during the cosmic dawn era. Among several proposed scenarios, the resonant conversion of dark radiation into photons is highly probable. Since the number density spectrum of the dark radiation can be several times higher than the CMB, a small resonant conversion probability can produce a large population of nonthermal radio photons. Future experiments such as SKA, DARE, SARAS 3, and REACH may provide confirmation or rule out the presence of such nonthermal photons, thereby shedding light on the BSM physics.

In this work, we consider ALPs as dark radiation that can resonantly convert to nonthermal radio photons in the presence of background magnetic fields. The existence of magnetic fields on large scales ($>1$~Mpc) suggests a primordial origin--- known as primordial magnetic fields (PMFs). Consequently, we consider that ALPs can resonantly convert to photons during the cosmic dawn era in the presence of PMFs. However, it is to be noted that PMFs can decay via the ambipolar-diffusion and the turbulent-decay and dissipate their energy into the IGM, resulting in additional heating of IGM that can reduce the absorption depth of the global 21-cm signal during the cosmic dawn. 
To obtain the $T_{21}$ signal, we calculate the thermal evolution of the IGM in the presence of both magnetic heating and resonant photon production. We consider a nearly scale-invariant spectrum of PMFs by fixing $n_B = -2.99$, since a mild increase in the spectral index can heat the IGM enough to make $T_g\gtrsim T_{\gamma}$. We varied the field strength within $0.1$–$0.55~\mathrm{nG}$ to explore the extent of IGM heating and its impact on $T_{21}$.

We then consider varying the absorptional amplitude in the range $(156-500)\,\rm mK$, depicting a wide range of possible scenarios from $\Lambda\rm CDM$ to the anomalous excess radio background during the cosmic dawn era. Lastly, in Fig. \eqref{fig:constrain} we present upper bounds on $g_{a\gamma}$ for ALP mass $m_a$ derived in the presence of magnetic heating for different 21-cm signals. We find that, for $B_n= 0.1\,\rm nG$, $g_{a\gamma}B_n\lesssim (3.6\times 10^{-4}-3\times 10^{-3})$ is required to recover $T_{21}=-156\,\rm mK$, while a deeper absorption of $-500$ mK pushes the upper bound to $g_{a\gamma}B_n\lesssim (6.5\times 10^{-4}-5.7\times 10^{-3})$.

In conclusion, unlike earlier studies such as article \cite{Moroi:2018vci}, which did not consider magnetic heating, our analysis incorporates both magnetic energy dissipation and resonant photon production. By incorporating magnetic heating of the IGM gas, we derive physically motivated bounds on $g_{a\gamma}$ that are consistent with the parameter space allowed by previous works, while extending the analysis to a broader and more realistic thermal history. Moreover, $g_{a\gamma}$ is $\propto B_n^{-1}$ and $\propto |T_{21}|$, therefore a balance between $B_n$ and $T_{21}$ is necessary to derive upper bounds. Our results demonstrate that resonant ALP-photon conversion remains a viable explanation for the excess radio background, provided the heating effect of primordial magnetic fields is carefully accounted for, thereby highlighting the need for upcoming 21-cm experiments to constrain or rule out such nonthermal photon excesses.

\bibliographystyle{ws-ijmpd}
\bibliography{sample}

\end{document}